\documentclass[reqno,centertags,11pt]{amsart}
\usepackage{amsmath,amsthm,amscd,amssymb} \usepackage{latexsym}
\usepackage{graphicx}
\usepackage{color}

 \newcommand{\N}{{\mathbb{N}}}
 \newcommand{\R}{{\mathbb{R}}}

 \newcommand{\Z}{{\mathbb{Z}}}

\newcommand{\bka}{\bk}
\renewcommand{\S}{ \mathbb S^{d-1}}

\newcommand {\bx}{\mathbf x}
\newcommand {\bk}{\mathbf k}
\newcommand {\bm}{\mathbf m}

\newcommand {\boldom}{\boldsymbol\omega}
\newcommand {\bn}{\boldsymbol n}
\newcommand {\be}{\boldsymbol e}
\newcommand{\br}{\boldsymbol r}

 \newcommand{\wti}{\widetilde  }

\newcommand{\hatt}{\widehat} 
\newcommand{\beq}{\begin{equation}}
\newcommand{\eeq}{\end{equation}}
\newcommand{\bdm}{\begin{displaymath}}
\newcommand{\edm}{\end{displaymath}} \newcommand{\ba}{\begin{align}}
\newcommand{\ea}{\end{align}} \newcommand{\bpf}{\begin{proof}}
\newcommand{\epf}{\end{proof}}

\newcommand{\la}{\langle} \newcommand{\ra}{\rangle}

 \allowdisplaybreaks

\newcommand{\mathcalB}{\mathcal{B}}

\newcommand{\mathcalG}{\mathcal{G}}

\newcommand{\mathcalS}{{\mathcal{S}}}

\newcommand{\meas}{{\mathrm {meas}}}

\newtheorem{thm}{Theorem} \newtheorem{prop}[thm]{Proposition}
\newtheorem{lem}[thm]{Lemma} 
\newtheorem{cor}[thm]{Corollary}

\theoremstyle{definition}

 \newtheorem{remark}[thm]{Remark}

\newcounter{theoremi}[thm] 

\numberwithin{thm}{section} \numberwithin{equation}{section}

\begin{document}

\title[Ballistic Transport]{Ballistic Transport for 
Schr\"odinger Operators with  Quasi-periodic Potentials}
\author[Y.~Karpeshina, L.~Parnovski, R.~Shterenberg]{Yulia Karpeshina, Leonid Parnovski and Roman Shterenberg}

\address{Department of Mathematics, University of Alabama at Birmingham, University Hall, Room 4005,
1402 10th Avenue South,
Birmingham AL 35294-1241, USA}
\email{karpeshi@uab.edu}%


\address{Department of Mathematics, University College London,
Gower Street, London WC1E 6BT, UK}
\email{leonid@math.ucl.ac.uk}

\address{Department of Mathematics, University of Alabama at Birmingham, University Hall, Room 4005,
1402 10th Avenue South,
Birmingham AL 35294-1241, USA}
\email{shterenb@math.uab.edu}



\date{\today}

\begin{abstract}
We prove the existence of ballistic transport for a 
Schr\"odinger operator with a generic quasi-periodic potential in any dimension $d>1$. \end{abstract}

\maketitle


\section{Introduction}

\subsection{Prior results on ballistic transport} \label{prior}

It is well known that the spectral and dynamical properties of Schr\"odinger operators $H=-\Delta+V$ acting in ${\mathcal H} = L^2(\R^d)$ are related. A general correspondence of this kind is given by the RAGE theorem, e.g.\ \cite{ReedSimon}. Stated briefly, it says that solutions $\Psi(\cdot,t) = e^{-iHt} \Psi_0$ of the time-dependent Schr\"odinger equation are `bound states' if the spectral measure $\mu_{\Psi_0}$ of the initial state $\Psi_0$ is pure point, while $\Psi(\cdot,t)$ is a `scattering state' if $\mu_{\Psi_0}$ is (absolutely) continuous. However, knowing the spectral type is not sufficient to quantify transport properties more precisely, for example in terms of diffusion exponents $\beta$. These exponents, if they exist, characterize how time-averaged $m$-moments
\begin{equation} \label{moments}
\langle \langle X_{\Psi_0}^m \rangle \rangle_T := \frac{2}{T} \int_0^{\infty} \exp\left(-\frac{2t}{T}\right) \| X^{m/2} \Psi(\cdot,t)\|^2_{\mathcal H} \,dt,\ \ \ \ \ m>0
\end{equation}
of the position operator $X$ grow as a power $T^{m\beta}$ of time $T$, where $(Xu)(x): = |x|u(x)$ ($x\in\R^d$ and $m$ is a positive real number).  The special cases $\beta = 1$, $\beta=1/2$ and $\beta=0$ are interpreted as ballistic transport, diffusive transport, and dynamical localization, respectively.

In general, due to the possibility of fast travelling small tails, $\beta$ may depend on $m$. In this paper, we will restrict our attention to the most frequently considered case of the second moment $m=2$. The ballistic upper bound
\begin{equation} \label{genballistic}
\| X \Psi(\cdot,t) \|^2_{\mathcal H}  \le C_1(\Psi_0) T^2 + C_2(\Psi_0),
\end{equation}
and thus also its averaged version $\langle \langle X_{\Psi_0}^2 \rangle \rangle_T \le C_1(\Psi_0) T^2 + C_2(\Psi_0)$, is known to hold for general potentials $V$ with relative $\Delta$-bound less than one (in particular all bounded potentials) and initial states
\begin{equation} \label{RScond}
\Psi_0 \in \mathcal{S}_1:= \{f\in L^2(\R^d): |x|f \in L^2(\R^d),\, |\nabla f| \in L^2(\R^d)\},
\end{equation}
see \cite{RaSi}.  As most authors, we will work with the Abel mean used in \eqref{moments}, but note that the existence of a ballistic upper bound can be used to show that Abel means and Cesaro means $T^{-1} \int_0^T \ldots\,dt$ lead to the same diffusion exponents (see for example Theorem~2.20 in \cite{DT2010}).

In the late 1980s and 1990s methods were developed which led to more concrete bounds on diffusion exponents by also taking fractal dimensions of the associated spectral measures into account and showing that this gives lower transport bounds. In particular, again for the special case of the second moment, the Guarneri-Combes theorem \cite{G,G2,C,L} says that
\begin{equation} \label{GuarneriCombes}
\langle \langle X_{\Psi_0}^2 \rangle \rangle_T \ge C_{\Psi_0} T^{2\alpha/d}.
\end{equation}
for initial states $\Psi_0$ with uniformly $\alpha$-H\"older continuous spectral measure (and satisfying an additional energy bound in the continuum case \cite{C}). In dimension $d=1$ this says that states with an absolutely continuous spectral measure ($\alpha=1$) also will have ballistic transport (as by \eqref{genballistic} the transport can not be faster than ballistic). In particular, this means that in cases where the spectra of one-dimensional Schr\"odinger operators with limit or quasi-periodic potentials were found to have an a.c.\ component, e.g.\ \cite{AS,CD,DS,El,MC,MP,PF}, one also gets ballistic transport.

The bound (\ref{GuarneriCombes}) does not suffice to conclude ballistic transport from the existence of a.c.\ spectrum in dimension $d\ge 2$. In fact, examples of Schr\"odinger operators with absolutely continuous spectrum, but slower than ballistic transport have been found: A two-dimensional  `jelly-roll' example with a.c.\ spectrum and diffusive transport is discussed in \cite{KL}, while \cite{BSB} provides examples of separable potentials in dimension $d\ge 3$ with a.c.\ spectrum and sub-diffusive transport.

In general, growth properties of generalized eigenfunctions should be used in addition to spectral information for a more complete characterization of the dynamics. General relations between eigenfunction growth and spectral type as well as dynamics were found in  \cite{KL}. A series of works studied one-dimensional models with $\alpha<1$ and related the dynamics to transfer matrix bounds, e.g.\ \cite{DLS,DT,DT2,DT3,GKT,JSBS,Tch}. In particular, these methods can establish lower transport bounds in models with sub-ballistic transport, such as the Fibonacci Hamiltonian and the random dimer model.

Until recently, much less has been done for $d\ge 2$.  Ballistic lower bounds and thus the existence of waves propagating at non-zero velocity were known only for $V=0$, where this is classical, e.g.\ \cite{ReedSimon}, and for periodic potentials \cite{AK}. Scattering theoretic methods show that this extends to potentials of sufficiently rapid decay, or sufficiently rapidly decaying perturbations of periodic potentials. 
In \cite{KLSS} two results on ballistic lower bounds in dimension $d=2$ were obtained, one for  limit-periodic and one for  quasi-periodic potentials. Our goal here is to generalize these result to any dimension $d\geq 2$ and a generic quasi-periodic potential.

We have already proved, in \cite{KPS}, that generic quasi-periodic potentials have absolutely continuous spectrum for high energies. Here, we will combine results obtained in \cite{KPS} (in particular, the properties of the generalised eigenfunctions constructed there) with the methods of \cite{KLSS} to prove the existence of 
 ballistic transport.


\subsection{The Main Result}

We study the initial value problem
 \begin{equation}\label{IVP}
   i \frac{\partial \Psi}{\partial t}=H\Psi, \ \ \ \Psi (\bx,0)=\Psi _0(\bx)
    \end{equation}
for  multidimensional Schr\"odinger operator $H$ acting on $L^2(\R^d)$, $d\ge 2$, defined in the following way. 
Let $\boldom_1,\dots,\boldom_l\in\R^d$, $l>d$, be a collection of vectors that we will call {\it the basic frequencies}. It will be convenient to form a `vector' out of the basic frequencies: 
$\vec\boldom:=(\boldom_1,\dots,\boldom_l)$. 
We consider the operator 
\begin{equation}\label{limper}
H:=-\Delta+V,
\end{equation}
where
\begin{equation}\label{V_q=0}
V:=\sum_{|\bn|\le Q} V_{\bn}\be_{\bn\vec\boldom}.
\end{equation}
The last sum is finite and taken over all vectors $\bn=(n_1,\dots,n_l)\in\Z^l$ with 
\begin{equation}\label{l_infinity}
|\bn|:=\max_{j=1,...,l}|n_j|<Q, \ \ \ Q\in\N.
\end{equation}
We have also denoted 
\begin{equation}\label{be}
\be_{\boldsymbol\theta}(\bx):=e^{i\left<\boldsymbol\theta,\bx\right>}, \ \ \boldsymbol\theta,\bx\in\R^d 
\end{equation}
and
\begin{equation}
\bn\vec\boldom:=\sum_{j=1}^l n_j\boldom_j\in\R^d; 
\end{equation}
these vectors  $\bn\vec\boldom$ being called {\it the frequencies}. 
For convenience and without loss of generality, we assume that the basic frequencies 
$\boldom_j\in [-1/2,1/2]^d$ and thus $\vec\boldom\in [-1/2,1/2]^{dl}$ (so that the Lebesgue measure of this set is one; obviously, we can always achieve this by rescaling). We assume the frequencies  $\boldom_1,..., \boldom_l$    are linearly independent over rationals. We also assume  $V_{-\bn}={\bar V_{\bn}}$. Clearly, $V$ is real valued.

Consider the evolution equation \eqref{IVP} for operators $H$ described above. Clearly, the ballistic upper bound of \cite{RaSi} can be applied and we have \eqref{genballistic} for initial conditions $\Psi_0$ satisfying (\ref{RScond}). We prove that for these  operators  there are  corresponding {\it ballistic lower bounds} for a large class of initial conditions. To formulate our main result, we use the infinite-dimensional spectral projection $E_{\infty}$ for $H$ whose construction is described in Section~\ref{SpecPropH} below.


\begin{thm}\label{Thm1}
For any given set of Fourier coefficients $\{V_{\bn}\}$, $V_{-\bn}={\bar V_{\bn}}$, $|\bn|\le Q$, $Q\in\N$, there exists a subset $\Omega_*=\Omega_*(\{V_{\bn}\})\subset [-1/2,1/2]^{dl}$ of basic frequencies with $\hbox{meas}\,(\Omega_*)=1$ such that for any $\vec\boldom\in\Omega_*$ there is an infinite-dimensional projection  $E_{\infty}=E_{\infty}(V)$ in $L^2(\R^d)$ (described in Section~\ref{SpecPropH}) with the following property: For any
\begin{equation} \label{nontriv}
\Psi _0\in {\mathcal C_0^\infty} \quad \mbox{with} \quad E_{\infty }\Psi _0 \neq 0
\end{equation}
there are constants $c_1 = c_1(\Psi_0)>0$ and $T_0 = T_0(\Psi _0)$ such that the solution $\Psi (\bx,t)$ of \eqref{IVP} satisfies the estimate
    \begin{equation}\label{ball-1}
    \frac{2}{T}\int _0^\infty e^{-2t/T} \big\|X \Psi (\cdot,t)\big \|^2_{L^2(\R^d)} dt >c_1 T^2
    \end{equation}
for all $T>T_0$.
\end{thm}

\begin{remark}
The set $\Omega_*$ in the formulation of the Theorem is implicit. More specifically, it is the very same set for which the results from \cite{KPS} are valid. In particular, the frequencies in this set satisfy Strong Diophantine Condition (see \cite{KPS} for more details). In what follows, we will assume that the potential $V$ is fixed and corresponding  frequencies belong to $\Omega_*$. We also remark that the notation in this paper, while following in most symbols the notation of \cite{KLSS} and \cite{KPS}, sometimes differs slightly from it. For example, the projection $E_{\infty}$ is denoted by $E^{(\infty)}$ in \cite{KPS}, etc.   
\end{remark}

In Section 2 we show that $E_{\infty}$  is close in norm to ${\mathcal F}^*\chi \left(\mathcal{G}_{\infty}\right){\mathcal F}$, where ${\mathcal F}$ is the Fourier transform and $\chi \left(\mathcal{G}_{\infty}\right) $ the characteristic function of a set $\mathcal{G}_{\infty}$, which has asymptotically full measure in $\R^d$, see (\ref{full}) and (\ref{March21a}).

As already remarked in Section~\ref{prior}, due to the validity of the ballistic upper bound (\ref{genballistic}) for all initial conditions $\Psi_0 \in \mathcal C_0^{\infty} \subset \mathcal{S}_1$, Theorem~\ref{Thm1} remains true if the Abel means are replaced by Cesaro means.

Theorem~\ref{Thm1} will be proven in two steps. First we will show

\begin{prop}\label{Prop2}
If $\Psi _0\in E_{\infty }\mathcal C_0^\infty$, $\Psi _0\neq 0$, $E_\infty$ being defined as in \cite{KPS},
then the solution $\Psi (\bx,t)$ of \eqref{IVP} satisfies the ballistic lower bound (\ref{ball-1}).
\end{prop}

Note that Proposition~\ref{Prop2} differs from Theorem~\ref{Thm1} by the fact that the initial condition $\Psi_0$ for which the ballistic lower bound is concluded is in the image of $\mathcal C_0^{\infty}$ under the projection $E_{\infty}$ (but that $\Psi_0$ itself is not in $\mathcal C_0^\infty$ here). This proposition takes the role of our core technical result, i.e.\ most of the technical work towards proving Theorem~\ref{Thm1} will go into the proof of the proposition. Theorem~\ref{Thm1} gives a more explicit description of initial conditions for which ballistic transport can be established. In fact, one easily combines Theorem~\ref{Thm1} with the ballistic upper bound (\ref{genballistic}) to get ballistic transport in form of a two-sided bound for many initial conditions:

\begin{cor} \label{Cor}
There is an $L^2(\R^d)$-dense and relatively open subset $\mathcal D$ of $\mathcal C_0^\infty(\R^d)$ such that for every $\Psi_0 \in {\mathcal D}$ there are constants $0<c_1 \le C_1 <\infty$ such that the ballistic upper bound (\ref{genballistic}) and the ballistic lower bound (\ref{ball-1}) holds for $T>T_0(\Psi _0)$.
\end{cor}

This follows by an elementary argument using only that $E_{\infty}$ is not the zero projection and $\mathcal C_0^{\infty}(\R^d)$ is dense in $L^2(\R^d)$ (and that $\mathcal C_0^{\infty}(\R^d)$ functions also satisfy (\ref{RScond})).

It is certainly desirable to go beyond this corollary and to more explicitly characterize classes of initial conditions for which (\ref{nontriv}) holds. This requires to much better describe and exploit the nature of the projection $E_{\infty}$.  While we believe that $E_{\infty }\Psi _0 \neq 0$ for any non-zero $\Psi _0\in {\mathcal C_0^\infty}(\R^d)$, we do not have a proof of this. We will return to this question later, see Remark~\ref{smoothEFE}, where we will more explicitly construct initial conditions which lead to both upper and lower ballistic transport bounds. These will have the form of suitably regularized generalized eigenfunction expansions.

As mentioned above, the proof of Theorem~\ref{Thm1} is very similar to the two-dimensional proof \cite{KLSS}. We just need to use the recent results from \cite{KPS} instead of 
those in \cite{KS1}. In what follows, we present the main steps in the proof and explain the changes we need to make in the proof due to the increase in dimension. 

\subsection*{Acknowledgments} The authors would like to dedicate this paper to the memory of Jean Bourgain.

The results were partially obtained during the programme Periodic and Ergodic Spectral Problems in January--June 2015, supported by EPSRC Grant
EP/K032208/1. YK is grateful to Mittag-Leffler Institute for
their support and hospitality (April, 2019).
The research of YK and RS was partially supported by NSF grant DMS--1814664.
The research of LP was partially supported by EPSRC grants EP/J016829/1 and EP/P024793/1.


\section{Spectral Properties of the Operator $H$} \label{SpecPropH}

Our proofs of Proposition~\ref{Prop2} and Theorem~\ref{Thm1} are based on the results and properties of quasi-periodic Schr\"odinger operators derived in the paper \cite{KPS}. While that work has derived, in particular, the existence of an absolutely continuous component of the spectrum, we will show now how the bounds obtained in \cite{KPS} for the spectral projections can be used to  prove the existence of ballistic transport. In this section we give a thorough discussion of the results and some of the methods from \cite{KPS}. In particular, we give a detailed construction of the spectral projection $E_{\infty}$ used in our main results. Unless stated otherwise, all statements in this section have been proved in \cite{KPS}.

\subsection{Prior results.} For any given set of Fourier coefficients $\{V_{\bn}\}$, $V_{-\bn}={\bar V_{\bn}}$, $|\bn|\le Q$, $Q\in\N$, there exists a subset $\Omega_*=\Omega_*(\{V_{\bn}\})\subset [-1/2,1/2]^{dl}$ of basic frequencies with $\hbox{meas}\,(\Omega_*)=1$ such that for any $\vec\boldom\in\Omega_*$ the following statements hold, for sufficiently small positive number $\sigma$, depending on $V$, $l$ and $d$ only. 

 
    \begin{enumerate}
    \item The spectrum of the operator (\ref{limper})
    contains a semi-axis. 
    
    \item There are generalized eigenfunctions $U_{\infty }(\bk, \bx)$,
    corresponding to the semi-axis, which are  close to the unperturbed exponentials. More precisely, 
    for every $\bk $ in an extensive (in the sense of (\ref{full}) below) subset $\mathcal{G} _{\infty }$ of $\R^d$  there is
    a solution $U_{\infty }(\bk, \bx)$ of the  equation
    $$HU _{\infty }=\lambda _{\infty }U_{\infty }$$ 
    that satisfies the following properties:
        \begin{equation} \label{aplane}
        U_{\infty }(\bk, \bx)
        =e^{i\langle \bk, \bx \rangle}\left(1+u_{\infty}(\bk, \bx)\right),
        \end{equation}
        \begin{equation}\label{aplane1}
        \|u_{\infty}\|_{L^\infty(\R^d)}=_{|\bk| \to \infty}O(|\bk|^{-\gamma _1}),\ \ \ \gamma _1=1-\sigma>0,
        \end{equation}
    where $u_{\infty}(\bk, \bx)$ is a quasi-periodic function:
    \begin{equation}\label{u}
u_{\infty}(\bk, \bx):=\sum_{\br \in \Z^l} c_{\br}(\bk)\be_{\br\vec \boldom}(\bx),
\end{equation}
the series converging in $L_{\infty }(\R^d)$.
    The  eigenvalue $\lambda _{\infty }(\bk)$ corresponding to
    $U_{\infty }(\bk, \bx)$ is close to $|\bk|^{2}$:
        \begin{equation}
        \lambda _{\infty }(\bk)=_{|\bk| \to \infty}|\bk|^{2}+
        O(|\bk|^{-\gamma _2}),\ \ \ \gamma _2=2-\sigma>0. \label{16a}
        \end{equation}
    The ``non-resonant" set $\mathcal{G} _{\infty }$ of the vectors $\bk$, for which (\ref{aplane}) to (\ref{16a}) hold, 
can be expressed as 
    $\mathcal{G} _{\infty }=\cap _{n=1}^{\infty }\mathcal{G}_n$,
    where $\{\mathcal{G} _n\}_{n=1}^{\infty}$ is a decreasing sequence of sets in $\R^d$. Each $\mathcal{G} _n$ has a finite number of holes in each bounded region. Typically, as $n$ increases,    
    more holes of smaller sizes appear in the intersection. As a result, the overall intersection $\mathcal{G} _{\infty }$ is, typically, a Cantor type set (i.e., it has empty interior). 
    This set 
    satisfies the estimate:
       \begin{equation}\label{full}
       \frac{\meas\left(\mathcal{G} _{\infty}\cap B_R\right)}{\meas( B_R)}
       =_{ R\to \infty }1+O( R^{-c\sigma}),\quad \sigma >0, \ \ \ c=c(l,d,\vec \boldom ),  
       \end{equation} 
    where $B_R$ is the ball of radius $R$ centred at the origin.

    \item The set $\mathcal{D}_{\infty}(\lambda)$, defined as a level (isoenergetic) set for
    $\lambda _{\infty }(\bk)$,
    $$ {\mathcal D} _{\infty}(\lambda)=\left\{ \bk \in \mathcal{G} _{\infty } :\lambda _{\infty }(\bk)=\lambda\right\},$$ is a slightly distorted sphere, typically with 
    infinite number of holes. It can be described by  the formula:
        \begin{equation} \label{D}
        {\mathcal D}_{\infty}(\lambda)=\{\bk:\bk=\varkappa _{\infty}(\lambda, \vec{\nu})\vec{\nu},
        \ \vec{\nu} \in {\mathcal B}_{\infty}(\lambda)\},
        \end{equation}
    where ${\mathcal B}_{\infty }(\lambda )$ is a subset of the unit
    sphere  $\S$. The set ${\mathcal B}_{\infty }(\lambda )$ can be
    interpreted as the set of possible directions of propagation for the
    almost plane waves (\ref{aplane}).  The set ${\mathcal B}_{\infty
    }(\lambda )$ typically has a Cantor type structure and has an asymptotically full
    measure on $\S$ as $\lambda \to \infty $:
        \begin{equation}\label{B}
        \hbox{meas}\,\left({\mathcal B}_{\infty}(\lambda )\right)
        =_{\lambda \to \infty}\hbox{meas}\,(\S) +O\left(\lambda^{-c\sigma}\right). 
        \end{equation}
    The value $\varkappa _{\infty }(\lambda ,\vec \nu )$ in (\ref{D}) is
    the ``radius" of ${\mathcal D}_{\infty}(\lambda)$ in a direction
    $\vec \nu $. The function $\varkappa _{\infty }(\lambda ,\vec \nu)-\lambda^{1/2}$ describes the deviation of ${\mathcal D}_{\infty}(\lambda)$ from the perfect sphere of radius
    $\lambda^{1/2}$. It is proven that the deviation is asymptotically
    small, uniformly in $\vec \nu \in {\mathcal B}_{\infty}(\lambda)$:
        \begin{equation}\label{h}
        \varkappa _{\infty }(\lambda ,\vec \nu)
        =_{\lambda \to \infty} \lambda^{1/2}+O\left(\lambda^{-\gamma _3}\right), \ \ \ \gamma _3=(3-\sigma)/2>0.
        \end{equation}

    \item The part of the spectrum corresponding to $\{U_{\infty }(\bk, \bx)\}_{\bk}$ is absolutely continuous.

\end{enumerate}

\begin{remark}
While parameter $\sigma$ can be chosen arbitrary small, all constants in $O(\cdot)$ depend on $\sigma$. 
For the purposes of this paper, we will not need to impose any additional assumptions on $\sigma$ on top of those assumed in \cite{KPS} (in particular, $\sigma<(100d)^{-1}$).
\end{remark}

\subsection{Description of the method.} \label{methods} To prove the results formulated in previous sub-section, in \cite{KPS} we have considered the sequence of operators $H_n=H_n(\bk)$, each being restriction of the operator $H$ onto the linear subspace of $\Z^l$ spanned by the exponentials $\be_{\bk+\bn\vec\boldom}$, $|\bn|\leq |\bk|^{r_n}$. Here, $r_n$ is a super exponentially growing sequence of numbers of the form
    
Each operator $H_n$, $n\geq 0$, is considered as a perturbation of
the previous operator $H_{n-1}$ ($H_{-1}=-\Delta$). For every operator
$H_n$, there is one eigenvalue located sufficiently far (at least $\sim |\bk|^{-r_n}$ away) from the rest of the spectrum of $H_n$. Corresponding eigenvector is close to the unperturbed exponential. More precisely,
for every $\bk $ in a certain subset $\mathcal{G} _n$ of $\R^d$, there
is a solution $U_{n}(\bk, \bx)$ of the differential equation
$H_n U_n=\lambda _nU_n$ that satisfies the following asymptotic formula:
    \begin{equation}\label{na}
    U_n (\bk, \bx)
    =e^{i\langle \bk, \bx \rangle}\left(1+ u_{n}(\bk, \vec
    x)\right),\ \ \
    \| u_{n}\|_{L^{\infty }(\R^d)}\underset{|\bk| \to \infty}{=}O(|\bk|^{-\gamma _1}),
    \end{equation}
where $ u_{n}(\bk, \cdot)$ is quasi-periodic, a finite combination of $\be_{\br\vec \boldom}(\bx)$:
 
    \begin{equation}\label{un}
u_{n}(\bk, \bx):=\sum_{\br \in \Z^l, |\br|<M_n} c_{\br}^{(n)}(\bk)\be_{\br\vec \boldom}(\bx),\ \ \ M_n\to \infty  \ \mbox{as}\ n\to \infty.
\end{equation}

The corresponding eigenvalue $\lambda_n (\bk)$ is close to $|\vec
k|^{2}$:
    \begin{equation}\label{back}
    \lambda_n(\bk)=_{|\bk| \to \infty}|\bk|^{2}+
    O\left(|\bk|^{-\gamma _2}\right).
    \end{equation}
The non-resonant set $\mathcal{G} _{n}$ for which (\ref{back}) holds,
is proven to be extensive in $\R^d$:
    \begin{equation}\label{16b}
    \frac{\meas(\mathcal{G} _{n}\cap
    B_R )}{\meas( B_R)}=_{R\to \infty }1+O(R^{-\sigma}).
    \end{equation}
The estimates (\ref{na}) -- (\ref{16b}) are uniform in $n$.
The set ${\mathcal D}_{n}(\lambda)$ is defined as the level
(isoenergetic) set for the non-resonant eigenvalue $\lambda_n(\vec
k)$: 
$$ {\mathcal D} _{n}(\lambda):=\left\{ \bk \in \mathcal{G}
_n:\lambda_n(\bk)=\lambda \right\}.$$ 
This set is a slightly distorted sphere with a finite number of holes; it can also be 
described by the formula:
     \begin{equation} {\mathcal D}_{n}(\lambda)=\{\bk:\bk=
    \varkappa_{n}(\lambda, \vec{\nu})\vec{\nu},
    \ \vec{\nu} \in {\mathcal B}_{n}(\lambda)\}, \label{Dn}
    \end{equation}
where ${\mathcal B}_{n}(\lambda )$ is a subset  of the unit sphere
$\S$. The set ${\mathcal B}_{n}(\lambda )$ can be interpreted as
the set of possible directions of propagation for  almost plane
waves (\ref{na}). The sequence of sets $\{{\mathcal
B}_n(\lambda)\}_{n=0}^{\infty }$  is decreasing,
since on each step more and more directions are excluded.  Each
${\mathcal B}_{n}(\lambda )$ has an asymptotically full measure on
$\S$ as $\lambda \to \infty $:
    \begin{equation}\label{Bn}
    \hbox{meas}\,\left({\mathcalB}_{n}(\lambda )\right)=_{\lambda \to \infty }\hbox{meas}\left(\S\right)
    +O\left(\lambda^{-\sigma/2}\right),
    \end{equation}
the estimate being uniform in $n$. The set ${\mathcal B}_{n}(\lambda)$ has
only a finite number of holes, however their number is growing with
$n$.  The value $\varkappa_{n}(\lambda ,\vec \nu
)-\lambda^{1/2}$ gives the deviation of ${\mathcal D}_{n}(\lambda)$
from the perfect sphere of  radius $\lambda^{1/2}$  in
direction $\vec \nu $. This deviation is
asymptotically small uniformly in $n$:
    \begin{equation}\label{hn}
    \varkappa_{n}(\lambda ,\vec \nu)
    =\lambda^{1/2}+O\left(\lambda^{-\gamma _3 }\right),\ \ \ \
    \frac{\partial \varkappa_{n}(\lambda ,\vec \nu)}{\partial \vec\varphi
    }=O\left(\lambda^{-\gamma _3 }\right),
    \end{equation}
$\vec\varphi $ being an angle variable associated with natural spherical coordinates (see \cite{KPS} for more details).

  More and more points are excluded from the
non-resonant sets $\mathcal{G} _n$ on each step.  Thus, $\{ \mathcal{G} _n
\}_{n=0}^{\infty }$ is a decreasing sequence of sets. The set
$\mathcal{G} _\infty $ is defined as the limit set: $\mathcal{G}
_\infty=\cap _{n=0}^{\infty }\mathcal{G} _n $. It has typically an infinite
number of holes in each bounded region, but nevertheless satisfies
the relation (\ref{full}). For every $\bk \in \mathcal{G} _\infty
$ and every $n$, there is a generalized eigenfunction of $H_n$
of the type  (\ref{na}). It is proven that  the sequence of $U
_n(\bk, \bx)$ has a limit in $L^{\infty }(\R^d)$ as $n\to
\infty$, when $\bk \in \mathcal{G} _\infty $. The function $U
_{\infty }(\bk, \bx) =\lim _{n\to \infty }U_n(\bk, \vec
x)$ is a generalized eigenfunction of $H$. It can be written in the
form (\ref{aplane})--(\ref{aplane1}). Naturally, the corresponding
eigenvalue $\lambda _{\infty }(\bk) $ is the limit of $\lambda_n(\bk )$ as $n \to \infty $.
Expansion with respect to the generalized eigenfunctions $\Psi
_{\infty }(\bk, \cdot)$, $\bk \in \mathcal{G} _\infty
$, will give a reducing subspace for $H$, with the corresponding spectral resolution arising as the limit of spectral resolutions for 
 operators $H_n$.

To study them, one needs properties of the limit ${\mathcal B}_{\infty}(\lambda)$ of ${\mathcal
B}_n(\lambda)$:
    $${\mathcal B}_{\infty}(\lambda)=\bigcap_{n=0}^{\infty} {\mathcal
B}_n(\lambda),\ \ \ {\mathcal B}_n(\lambda) \subset {\mathcal B}_{n-1}(\lambda).$$
 This set has asymptotically full measure, as (\ref{B}) follows from (\ref{Bn}).
The sequence $\varkappa _n(\lambda ,\vec \nu )$, $n=0,1,2,... $,
describing the isoenergetic sets $\mathcal{D}_n(\lambda)$,  quickly converges as $n\to \infty$.
Hence, ${\mathcal D}_{\infty}(\lambda)$ can be described as the
limit of ${\mathcal D}_n(\lambda)$ in the sense (\ref{D}), where
$\varkappa _{\infty}(\lambda, \vec{\nu})=\lim _{n \to \infty}
\varkappa _n(\lambda, \vec{\nu})$ for every $\vec{\nu} \in {\mathcal
B}_{\infty}(\lambda)$. The derivatives of the
functions $\varkappa _n(\lambda, \vec{\nu})$ (with respect to the
angle variable $\vec\varphi $) have a limit as $n\to
\infty $ for every $\vec{\nu} \in {\mathcal B}_{\infty}(\lambda)$.
We denote this limit by $\frac{\partial \varkappa_{\infty}(\lambda
,\vec \nu)}{\partial \vec\varphi }$. We also have 
    \begin{equation}\label{Dec9a}
    \frac{\partial \varkappa_{\infty}(\lambda ,\vec \nu)}{\partial
    \vec\varphi }=O\left(\lambda^{-\gamma _3 }\right).
    \end{equation}
Thus, the limit set ${\mathcal D}_{\infty}(\lambda)$ takes the form of
  a slightly distorted sphere with, possibly, 
infinite number of holes.

Let   $\mathcal{G}_n'$ be a bounded Lebesgue measurable subset of  $\mathcal{G}_n$.
We consider the spectral projection
$E_n\left(\mathcal{G}_n'\right)$ of $H_n$, corresponding to functions
$U_n(\bk ,\bx)$, $\bk \in \mathcal{G}_n'.$
By \cite{Ge}, $E_n\left( \mathcal{G}'_n\right): L^2(\R^d)\to
L^2(\R^d)$ can be represented by the formula:
    \begin{equation}\label{s}
    E_n\left( \mathcal{G}'_n\right)F=\frac{1}{(2\pi)^d}\int
    _{ \mathcal{G}'_n}\bigl( F,U_n(\vec
    k )\bigr) U_n(\vec
    k) \,d\bk
    \end{equation}
for any $F\in {\mathcal C_c}(\R^d)$, the space of continuous, compactly supported functions on $\R^d$. Here and below, $\bigl( \cdot ,\cdot \bigr)$
is the integral corresponding to the canonical scalar product in $L^2(\R^d)$, i.e.,
    $$
    \bigl( F,U_n(\bk )\bigr)=\int _{\R^d}F(\bx)\overline{U_n(\bk ,\bx)}\,d\bx.
    $$
The above formula can be rewritten in the form
    \begin{equation}\label{ST}
    E_n\left(\mathcal{G}'_n\right)=S_n\left(\mathcal{G}'_n\right)T_n \left(
    \mathcal{G}'_n\right),
    \end{equation}
    $$T_n:{\mathcal C_c}(\R^d) \to L^2\left(  \mathcal{G}'_n\right), \ \
    \ \ S_n:L^\infty\left( \mathcal{G}'_n\right)\to L^2(\R^d),$$
    \begin{equation}\label{eq2}
    (T_nF)(\bk): =\frac{1}{(2\pi)^{d/2} }\bigl( F,U_n(\vec
    k )\bigr) \mbox{\ \ for any $F\in {\mathcal C_c}(\R^d)$},
    \end{equation}
$T_nF$ being in $L^{\infty }\left(  \mathcal{G}'_n\right)$, and
    \begin{equation}\label{ev}
    (S_nf)(\bx): = \frac{1}{(2\pi)^{d/2} }\int _{  \mathcal{G}'_n}f (\bk)U_n(\vec
    k ,\bx)\,d\bk  \mbox{\ \ for any $f \in L^{\infty }\left(
    \mathcal{G}'_n\right)$.}
    \end{equation}
By \cite{Ge},
 \beq\label{eq:T_n bound}
 \|T_nF\|_{L^2\left( \mathcal{G}'_n\right)}\leq \|F\|_{L^2(\R^d)}
 \eeq
and
 \beq\label{eq:S_n bound}
 \|S_nf \|_{L^2(\R^d)}\leq \|f \|_{L^2\left(\mathcal{G}'_n\right)}.
 \eeq
Hence, $T_n$ and $S_n$ can be extended by continuity from ${\mathcal C_c}(\R^d)$ and $L^{\infty }\left(  \mathcal{G}'_n\right)$ to $L^2(\R^d)$ and $L^2\left( \mathcal{G}'_n\right)$,
respectively. Obviously, $T_n^{*}=S_n$. Thus, the operator $E_n\left( \mathcal{G}'_n\right)$ is described by (\ref{ST}) in the whole space $L^2(\R^d)$.

In what follows we will use these operators for the case where $\mathcal{G}'_n$ is given by
\begin{equation}
\mathcal{G}'_n=\mathcal{G}_{n, \lambda}:=\{ \bk \in {\mathcal{G}}_n:
\lambda_n(\bk) < \lambda\}. \label{d}
\end{equation}
for finite sufficiently large $\lambda$. This set is Lebesgue measurable since ${\mathcal{G}}_n $ is
open and $\lambda_n(\bk)$ is continuous on $
{\mathcal{G}}_n$.

Let
    \begin{equation}\label{dd}
    \mathcal{G}_{\infty, \lambda }=\left\{\bk \in
    \mathcal{G}_{\infty }: \lambda _{\infty }(\bk )<\lambda
    \right\}.
    \end{equation}
The function $\lambda _{\infty }(\bk )$ is a Lebesgue
measurable function, since it is a  pointwise limit of a sequence of
measurable functions. Hence, the set  $\mathcal{G}_{\infty, \lambda
}$ is measurable. The sets $\mathcal{G}_{n, \lambda
}$ and $\mathcal{G}_{\infty, \lambda
}$ are also bounded. The measure of the symmetric difference of the
two sets $\mathcal{G}_{\infty, \lambda }$ and $\mathcal{G}_{n,
\lambda}$ converges
 to zero as $n \to
\infty$, uniformly in $\lambda$ in every bounded interval:
    $$\lim _{n\to \infty }\meas(\mathcal{G}_{\infty, \lambda }\,\Delta\, \mathcal{G}_{n, \lambda
    })=0.$$

Next, we  consider the sequence of operators $S_n(\mathcal{G}_{\infty ,\lambda })$  given by (\ref{ev})
 and with ${\mathcal G}_n'=\mathcal{G}_{\infty , \lambda}$; 
    \begin{equation} \label{SnGDef}
    S_n(\mathcal{G}_{\infty , \lambda}):\ L^2(\mathcal{G}_{\infty , \lambda})\to L^2(\R^d).
    \end{equation}
This sequence 
has a limit $S_{\infty }(\mathcal{G}_{\infty , \lambda})$ in operator norm sense as $n\to\infty$, uniform in $\lambda$. 
Moreover, the estimate
 \begin{equation}
    \label{June5} \|S_{\infty }(\mathcal{G}_{\infty , \lambda})-S_{-1}(\mathcal{G}_{\infty , \lambda})\|<c\lambda _*^{-\gamma _1}
    \end{equation}
    holds for $\lambda > \lambda_*$, $c$ not depending on $\lambda , \lambda _*$.       Here we put $U_{-1}=e^{i\langle \bk,\bx\rangle}$ and define  $S_{-1}$ by  \eqref{ev}. The operator $S_{\infty }(\mathcal{G}_{\infty , \lambda})$ satisfies $\|S_{\infty }\|=1$ and can be described by the formula
    \begin{equation}\label{ev1}
    (S_{\infty }f) (\vec x)= \frac{1}{(2\pi)^{d/2} }\int _{\mathcal{G}_{\infty , \lambda}}f (\bk)\Psi _{\infty }(\vec
    k,\bx) \,d\bk
    \end{equation}
for any $f \in L^{\infty }\left( \mathcal{G}_{\infty , \lambda}\right)$.

    Similarly,  we consider the sequence of operators $T_n(\mathcal{G}_{\infty , \lambda})$ which are given by (\ref{eq2}) and act from $L^2(\R^d)$ to $L^2(\mathcal{G}_{\infty , \lambda})$.
Since $T_n=S_n^*$, the sequence $T_n(\mathcal{G}_{\infty ,\lambda})$  has a limit $T_{\infty}(\mathcal{G}_{\infty , \lambda})=S^*_{\infty }(\mathcal{G}_{\infty , \lambda})$ in operator norm sense. The operator $T_{\infty }(\mathcal{G}_{\infty , \lambda})$ satisfies $\|T_{\infty }\|\leq 1$ and can be described by the formula
$(T_{\infty }F)(\bk)=\frac{1}{(2\pi)^{d/2} }\bigl( F,\Psi _{\infty }(\bk)\bigr) $ for any $F\in {\mathcal C_c}(\R^d)$.
The convergence is uniform in $\lambda $
    and
     \begin{equation}
    \label{June5a} \|T_{\infty }(\mathcal{G}_{\infty , \lambda})-T_{-1}(\mathcal{G}_{\infty , \lambda})\|<c\lambda _*^{-\gamma _1}.
    \end{equation}


Spectral projections
$E_n(\mathcal{G}_{\infty , \lambda })$ converge in norm to
$E_{\infty }(\mathcal{G}_{\infty , \lambda })$ in $L^2(\R^d)$ as $n$
tends to infinity, since $T_n=S_n^*$. The operator $E_{\infty }(\mathcal{G}_{\infty ,
\lambda })$ is a spectral projection of $H$. It can be represented
in the form $E_{\infty }(\mathcal{G}_{\infty , \lambda })=S_\infty(\mathcal{G}_{\infty , \lambda })
T_{\infty }(\mathcal{G}_{\infty , \lambda })$.
For any $F\in
{\mathcal C_c}(\R^d)$ we have
    \begin{equation}\label{s1}
    E_{\infty }\left(\mathcal{G}_{\infty , \lambda }\right)F=\frac{1}{(2\pi)^d}\int
    _{ \mathcal{G}_{\infty , \lambda }}\bigl( F,\Psi _{\infty }(\vec
    k)\bigr) \Psi _{\infty }(\vec
    k ) \,d\bk ,
    \end{equation}
    \begin{equation}\label{s1uu}
    HE_{\infty }\left(\mathcal{G}_{\infty , \lambda }\right)F=\frac{1}{(2\pi)^d}\int
    _{ \mathcal{G}_{\infty , \lambda }}\lambda _{\infty }(\bk )
    \bigl( F,\Psi _{\infty }(\bk )\bigr) \Psi _{\infty }(\bk ) \,d\bk .
    \end{equation}
Since   $E_{\infty }$ is a projection, one has the Parseval formula
\begin{equation} \label{Parseval}
\| E_{\infty}(\mathcal{G}_{\infty,\lambda }) F\|^2 = \frac{1}{(2\pi)^d} \int_{\mathcal{G}_{\infty,\lambda}} |(F, \Psi_{\infty}(\bk)|^2\,d \bk.
\end{equation}
It is easy to see that
 \begin{equation}
    \label{June5b} \|E_{\infty }(\mathcal{G}_{\infty , \lambda})-S_{-1}T_{-1}(\mathcal{G}_{\infty , \lambda})\|<c\lambda _*^{-\gamma _1},
    \end{equation}
    \begin{equation}
    S_{-1}T_{-1}(\mathcal{G}_{\infty , \lambda})={\mathcal F}^*\chi (\mathcal{G}_{\infty , \lambda}){\mathcal F}. \label{June5c}
    \end{equation}

Projections $E_\infty
(\mathcal{G}_{\infty,\lambda })$ are increasing in $\lambda$ and have a strong limit
$E_\infty(\mathcal{G}_{\infty})$ as $\lambda $ goes to infinity.
Hence,  the operator $E_{\infty
}(\mathcal{G}_{\infty})$ is a projection. The projections
$E_\infty(\mathcal{G}_{\infty,\lambda })$, $\lambda\geq\lambda_*$, and
$E_\infty(\mathcal{G}_{\infty})$ reduce the operator $H$.
The family of projections
$E_\infty(\mathcal{G}_{\infty,\lambda} )$ is the resolution of the
identity of the operator $HE_\infty(\mathcal{G}_{\infty})$ acting in
$E_\infty(\mathcal{G}_{\infty})L^2(\R^d)$.  Let us denote $E_\infty:=E_\infty(\mathcal{G}_{\infty})$   and use
\begin{equation}
\|E_\infty  -{\mathcal F}^*\chi (\mathcal{G}_{\infty }){\mathcal F}\|<c\lambda _*^{-\gamma _1}.\label{March21a}
    \end{equation}
Obviously, the r.h.s.\ can be made arbitrarily small by an appropriate choice of $\mathcal{G}_{\infty }$.

The restriction of $H$ to the range of $E_{\infty}$ has purely absolutely continuous spectrum. In addition to the above mentioned convergence of the spectral projections of $H_n$ to those of $H$, uniform in $\lambda \ge \lambda_*$ for sufficiently large $\lambda_* = \lambda_*(V)$, this requires an analysis of the continuity properties of  the level sets
$\mathcal{D}_{\infty }(\lambda )$ with respect to $\lambda $.





\subsection{Extension of $\lambda_\infty(\bk)$  from $\mathcal G_\infty$  to $\R^d$} \label{extlambda}

Let $M$ be a large natural number; for the purposes of this paper, taking $M:=[3d/2 +6]$ would do. 
We want to extend the function  $\lambda_\infty(\bk)$ from $\mathcal G_\infty$ to $\R^d$, the result being a $\mathcal C^M(\R^d)$ function.
Note that the extended function is not going to be a generalised eigenvalue outside of $\mathcal G_\infty$.

 The first step is  representing $\lambda _{\infty }(\bk)-k^2$, $k:=|\bk|$, 
$\bk \in {{\mathcal G}}_{\infty }$,
in the form:
$$\lambda _{\infty }(\bk)-k^2=\lambda_0(\bk)-k^2+\sum _{n=1}^{\infty }\left( \lambda_{n}(\bk)-\lambda_{n-1}(\bk)\right). $$
Let $\bm=(m_1,...m_d)$ be a multi-index and put
 $D^{\bm}_{\bk} := \partial_1^{m_1} \cdots\partial_d^{m_d}$. We have (see \cite{KPS}, Lemma 11.3):
 \begin{equation}\label{eigenvalue-1}
 \left|  D^{\bm}_{\bk}\left(\lambda_0(\bk)-k^2\right)\right|<Ck^{-\gamma _2+\sigma|m|},\ \ \gamma _2=2-\sigma,
 \end{equation}
when $\bk $ is in the $k^{-\sigma}$-neighborhood of ${{\mathcal G}}_0\supset {{\mathcal G}}_{\infty }$ and
 \begin{equation}\label{derivative-eigenvalue-n*}
 \left| D^{\bm}_{\bk}\left(\lambda_{n}(\bk)-\lambda_{n-1}(\bk)\right) \right|< Ck^{-k^{r_{n-1}}+|m|k^{\sigma r_{n-1}}}
 \end{equation}
in the $k^{-k^{\sigma r_{n-1}}}$-neighbourhood of ${{\mathcal G}}_n$ for all $m$. Here, the constants depend only on $V$ and $m$.

Now, we introduce a function $\eta _0(\bk)\in{\mathcal C^\infty}(\R^d)$  with support in the (real) $k^{-\sigma}$-neighbourhood of ${{\mathcal G}}_0$,
satisfying $\eta _0=1$ on ${{\mathcal G}}_0$ and 
\begin{equation}\left|D^{m}\eta _0(\bk)\right|<C k^{\sigma|m|}. \label{eta0}
\end{equation}
This is possible since we can take a convolution of the characteristic function of the $\frac 12 k^{-\sigma}$-neighbourhood of ${{\mathcal G}}_0$ with  $w (2k^{\sigma} \bk)$,
where $\omega (\bk )$ is a non-negative ${\mathcal C_0^\infty}(\R^d)$-function with a support in the unit ball centred at the origin and integral one.
Similarly,  let $\eta _n(\bk)$, $n\geq 1$,  be a $\mathcal C^\infty$ function with support in the $k^{-k^{\sigma r_{n-1}}}$-neighbourhood of ${{\mathcal G}}_n$,
satisfying $\eta _n=1$ on ${{\mathcal G}}_n$ and
 \begin{equation}\label{derivative-eta-n}
 \left|D^{m}_{\bk}\eta _n (\bk)\right|\leq C k^{|m| k^{\sigma r_{n-1}}}.
 \end{equation}

Next, we extend $\lambda _{\infty }(\bk)-k^2$ from ${{\mathcal G}}_{\infty }$ to $\R^d$ using the formula
\begin{equation}\lambda _{\infty }(\bk)-k^2=(\lambda_0(\bk)-k^2)\eta _0(\bk)+\sum _{n=1}^{\infty }\left( \lambda_{n}(\bk)-\lambda_{n-1}(\bk)\right)\eta _{n}(\bk).\label{May31} \end{equation}
It follows from \eqref{eigenvalue-1} -- \eqref{derivative-eta-n}   that the series converges in $\mathcal C^M(\R^d)$. 
Taking into account  that $\sigma>0$ could be chosen arbitrary small (note that $\lambda_*$ increases and  $\mathcal G_{\infty}$ is getting smaller when $\sigma$ decreases) gives the following lemma:

\begin{lem}\label{lambda}
For every natural number $M$, there exists $\lambda_{*}(V,M)>0$ such that
 the function $\lambda _{\infty }(\bk)-k^2$ can be extended, as a
${\mathcal C^{M}}$ function, from ${{\mathcal G}}_{\infty}$ to $\R$ and it satisfies
\begin{equation}\label{eigenvalue}
\left|  D^{\bm} _{\bk}\left(\lambda _{\infty }(\bk)-k^2\right)\right|<C_M k^{-\gamma _2+\sigma|m|}, 
\end{equation}
for any $ m \in \N_0^d$ with $|m| \leq M<\sigma^{-1}$.
\end{lem}

 \subsection{Extension of $U_{\infty }(\bk,\bx)$ from $\mathcalG_\infty$ to $\R^d$} \label{extpsi}

We now define $U_{\infty }(\bk,\bx)$ for arbitrary $\bk\in\R^d$ by a formula analogous to \eqref{May31}:
\begin{equation} \label{May31a}
 U_{\infty }(\bk,\bx)-e^{i\la \bk, \bx \ra}
 =\left(\Psi_0(\bk,\bx)-e^{i\la \bk, \bx \ra}\right)\eta_0(\bk)
  + \sum_{n=1}^\infty \left(U_{n}(\bk,\bx)-U_{n-1}(\bk,\bx)\right)\eta_{n}(\bk),\end{equation}
  here $U_n$ are described by \eqref{na}, \eqref{un}, and (see \cite{KPS}, Lemma 11.3)
   \begin{equation}
  \begin{split}
\label{psi1}
   \|D_{\bk}^m\left(u^{(0)}-\be_{\bka}\right)\|_{L_{\infty }(\R^d)}<C k^{-\gamma_1+\sigma |m|},\ \ \gamma_1=1-\sigma,\\
 \end{split}
    \end{equation}
   \begin{equation}
    \begin{split} \label{Dec10}
    &\|D_{\bk}^m\left(u^{(n)}-u^{(n-1)}\right)\|_{L_{\infty}(\R^d)}
    < C k^{-k^{r_{n-1}}+|m| k^{\sigma r_{n-1}}}, 
\end{split}
    \end{equation}
 Thus, the series \eqref{May31a} is convergent in $L_{\infty}(\R^d)$. Using the last formula and \eqref{ev1}, we define $S_\infty(\wti{\mathcalG}_\infty )$ for any $\wti{\mathcalG}_\infty \supset {\mathcalG}_\infty $:

\begin{align} \label{Sinf} \left(S_\infty(\wti{\mathcalG}_\infty )f\right)(\bx)
  &:= \frac{1}{(2\pi)^{d/2}}\int_{\wti{\mathcalG}_\infty } f(\bk) U_{\infty }(\bk,\bx)\,d\bk.
   \end{align}
  It is easy to see that
\begin{equation} \label{June}
  S_\infty(\wti{\mathcalG}_\infty )
  = S_{-1}(\wti{\mathcalG}_\infty )+\sum_{n=0}^\infty \bigl(S_{n}(\wti{\mathcalG}_\infty )-S_{n-1}(\wti{\mathcalG}_\infty ) \bigr)\eta_{n},
\end{equation}
 where $S_{-1}(\wti{\mathcalG}_\infty )$ is defined by
 $$S_{-1}(\wti{\mathcalG}_\infty )f= \frac{1}{2\pi}\int _{\wti{\mathcalG}_\infty }f(\bk)e^{-i\la \bk, \bx \ra}d\bk,$$
 $\eta_{n}$ is multiplication by $\eta_{n}(\bk)$ and $S_n(\wti{\mathcalG}_\infty ) $ is given by \eqref{ev} with ${\mathcalG}_n'$ being the intersection of
 $\wti{\mathcalG}_\infty $ with the $k^{-k^{\sigma r_{n-1}}}$-neighborhood of ${{\mathcal G}}_n$ for $n\geq 1$ and  the $k^{-\sigma}$-neighborhood of ${{\mathcal G}}_0$ for $n=0$.

Similarly to \eqref{June5}, we show that
 \begin{equation}
  \|S_\infty(\wti{\mathcalG}_\infty )-S_{-1}(\wti{\mathcalG}_\infty )\|<c(V)\lambda_{*}^{-\gamma _1}.
  \label{2.36}
  \end{equation}
In what follows we assume that $\lambda_{*}$ is chosen so that, in particular, $c(V)\lambda_{*}^{-\gamma _1}\leq 1/2$ (in fact, this is already the case under conditions from \cite{KPS}). Clearly, $ \| S_{-1}(\wti{\mathcalG}_\infty )\|=1$. Thus we have
 \begin{equation}\label{S_infty}
 \| S_\infty(\wti{\mathcalG}_\infty )\|\leq 2.
 \end{equation}
Similarly, with $T_{-1}$ being the Fourier transform,
\begin{align}\label{T_infty}
  (T_\infty F)(\bk) & := \frac{1}{(2\pi)^{d/2}} (F(\cdot), U_{\infty }(\bk, \cdot)) \notag \\
  &= (T_0 F)(\bk)+ \sum_{n=0}^\infty \bigl((T_{n}-T_{n-1})F\bigr)(\bk)\eta_{n}(\bk).
  \end{align}
 .


 \begin{lem}\label{Ttransform}
 For any given $L\in\N$ there exists $\lambda_{*}(V,L)$ such that for any $F \in {\mathcal C^\infty_0}(\R^d)$, the function $T_\infty F$ as defined above is in ${\mathcal C^L}(\R^d)$. Moreover, if $0\leq j \leq L$ and $m \in \N_0^d,\ |m|\leq L$, then
  \begin{equation}\label{T_infty bound}
  \left| |\bk |^j D^{\bm} (T_\infty F) (\bk)\right| < C(L, F),
  \end{equation}
 for all $\bk \in \R^d$.
 \end{lem}
 We prove the lemma using \eqref{T_infty} and then \eqref{eq2} for each $T_n$. Integrating by parts $j$ times and considering \eqref{derivative-eta-n}, \eqref{psi1}, and \eqref{Dec10}, we arrive at \eqref{T_infty bound}.
 \begin{remark}
 For our needs $L=M=[3d/2+6]$ is sufficient, so we may assume that such $L$ and $M$ are fixed.
 \end{remark}

\section{Proofs of Proposition~\ref{Prop2} and Theorem \ref{Thm1}} \label{Prop2proof}

Let $\mathcalS:=T_\infty{\mathcal C_0^\infty}(\R^d)$, see \eqref{T_infty}. Let $\hatt{\Psi}_0 \in \mathcalS$. As shown in Lemma~\ref{Ttransform},
  then
 \begin{equation}
 \bigl| |\bk|^j D^{m}(\hatt{\Psi}_0)(\bk)| <C(j,m,\hatt{\Psi}_0)
 \end{equation}
 for any $\bk \in \R^d$.

 Now we define 
\beq\label{Def:Psi}
\Psi(\bx,t):= \frac{1}{(2\pi)^{d/2}}\int_{\mathcalG_\infty} U_{\infty }(\bk,\bx)e^{-it\lambda_\infty(\bk)}\hatt{\Psi}_0(\bk)\,d\bk ,
\eeq
then this function solves the initial value problem \eqref{IVP}, where
\beq\label{Def:Psi-0}\Psi _0 (\bx) =\frac{1}{(2\pi)^{d/2}}\int_{\mathcalG_\infty} U_{\infty }(\bk,\bx)\hatt{\Psi}_0(\bk)\,d\bk  \eeq
and $\Psi _0 (\bx) \in S_\infty \mathcal S=E_\infty\mathcal C_0^\infty$.  Obviously,  $S_\infty \mathcal S$ is dense in  $E_\infty L^2(\R^d)$.

The next step of the proof is replacing $\mathcalG_\infty$ by a small neighbourhood $\tilde \mathcalG_\infty $ and to estimate the resulting errors in the integrals. This is an important step, since  $\mathcalG_\infty$ is a closed Cantor-type set, while $\tilde \mathcalG_\infty $ is an open set.
Then we would like to integrate by parts in
the  integral over $\tilde \mathcalG_\infty $ with the purpose of obtaining \eqref{ball-1}; the fact that $\tilde \mathcalG_\infty $ is open being used for handling the boundary terms.

To get the lower bound \eqref{ball-1}, we first note that
$$ \|X\Psi\|^2_{L^2(\R^d)} \geq \|X\Psi\|^2_{L^2(B_R)} \geq \frac{1}{2}\|Xw\|^2_{L^2(B_R)}-\|X(\Psi-w)\|^2_{L^2(B_R)},$$
where $B_R$ is the open disc with radius $R$ centred at the origin, $R=c_0T$, $c_0$ to be chosen later, and $w(\bx,t)$ is an approximation of $\Psi $ when $\mathcalG_\infty$ is replaced by its small neighbourhood $\tilde \mathcalG_\infty $. Namely,
\beq\label{Def:w}
w(\bx,t):=\frac{1}{(2\pi)^{d/2}}\int_{\wti{\mathcalG}_\infty}U_{\infty }(\bk,\bx)e^{-it\lambda_\infty(\bk)}\hatt{\Psi}_0(\bk)\tilde\eta_\delta(\bk)\,d\bk,
\eeq
$\tilde\eta_\delta$ being a smooth cut-off function with support in a $\delta$-neighbourhood $\wti{\mathcalG}_\infty$ of $\mathcalG_\infty$ and $\tilde\eta_\delta=1$ on $\mathcalG_\infty$.
The parameter $\delta$ $(0<\delta<1)$ will be chosen later to be sufficiently small and depend only on $\hatt\Psi_0$.  We take $\tilde\eta_\delta$  to be a convolution of a function
$\omega (\bk/2\delta )$ with the characteristic function of the $\delta/2$-neighbourhood of $\mathcalG_\infty$, where $\omega$ is a smooth cut-off function defined in the previous section. Then, $\tilde\eta _\delta \in {\mathcal C_0^\infty}(\R^d)$,
\begin{equation}
0\leq \tilde\eta _\delta \leq 1,\  \tilde\eta _\delta(\bk)=1\; \mbox{when } \bk\in \mathcalG_\infty,\ \tilde\eta _\delta(\bk)=0\;  \mbox{when } \bk  \not \in \wti{\mathcalG}_\infty,\
\|D^{m}\tilde\eta _\delta \|_{L^{\infty }}<C_{m}\delta ^{-|m |}. \label{eta-delta}
\end{equation}

To prove \eqref{ball-1}, we will show that there exist a positive constant $c_1$ and  constants $c_2$ and $c_3$  such that
    \beq\label{Ineq:main}
    \frac{2}{T}\int _0^\infty e^{-2t/T} \big\|Xw(\cdot, t)\big \|^2_{L^2(B_R)} dt \geq 6c_1 T^2 - c_2 T - c_3,
    \eeq
as long as $c_0$ in the definition of $R$ exceeds a certain value depending only on $\hatt \Psi _0$. In formula  \eqref{Ineq:main}, the constant $c_1=c_1(\hatt \Psi_0)$ depends on $ \hatt \Psi_0$, but not $\delta $ or $c_0$, while the constants $c_2=c_2(\hatt \Psi_0,\delta )$ and $c_3=c_3(\hatt \Psi_0,\delta )$ depend on $ \hatt \Psi_0$ and $\delta $, but not $c_0$.

We also prove that
    \beq\label{Ineq:remainder}
    \frac{2}{T}\int _0^\infty e^{-2t/T} \big\|X(\Psi-w)(\cdot, t)\big \|^2_{L^2(B_R)} dt \leq \gamma(\delta, \hatt \Psi_0)c_0^2 T^2,
    \eeq
  $\gamma(\delta, \hatt \Psi_0)=o(1)$  as  $\delta\to 0$ uniformly in $c_0$.

The proofs of \eqref{Ineq:main} and \eqref{Ineq:remainder} are completely analogous to those from \cite{KLSS}. The only difference is in the estimate of the integral of the form

\begin{equation}\label{Hor}
\tilde \phi _1(\vec z t,t):=\frac{1}{(2\pi)^{d/2}} \int_{\wti{\mathcalG}_\infty \cap \{\bk \, :\, \left| \bk - \bk_0 \right| < 2 \} } e^{it\left(\la \bk,\vec z\ra
    -\lambda_\infty(\bk)\right)}g_3(\bk)\bigl(1-\hat\eta(\bk)\bigr)\, d\bk,\ \ \ \vec z:=\frac{\bx}{t},
    \end{equation}
    where $g_3(\bk):=\nabla \lambda_\infty(\bk) \hatt{\Psi}_0(\bk)\tilde\eta_\delta(\bk)$, $\hat\eta$
     is a smooth cut-off function satisfying
 $$
 \hat\eta(\bk)=
  \begin{cases}
   0, & \left| \bk -\bk_0 \right| \le 1 \\
   1, & \left| \bk -\bk_0 \right| \ge 2
  \end{cases}
$$
and 
$$
\bk_0 =  \bk_0(\vec z)=\frac{1}{2} \vec z + O(|\vec z|^{-\gamma_4}),\ \ \gamma _4>0,
 $$ 
 is the unique solution (see \eqref{May31} and Lemma~\ref{lambda}) of the equation for a stationary point
 $$
 \vec z- \nabla \lambda_\infty (\bk) =0, \ \ \ |\vec z|^2>\lambda_*.
 $$


As in \cite{KLSS}, we apply Theorem~7.7.5 in \cite{H1} but for arbitrary $d>1$. The number of derivatives required depends on the dimension ($M:=[3d/2+6]$ is enough). We have
\begin{equation}\label{tilde_phi_1}
\tilde \phi _1(\vec z t,t)=\frac{1}{(2i)^{d/2}} e^{it\left(\la \bk_0,\vec z\ra -\lambda_\infty(\bk_0)\right)}
 \left(1+O(|\vec z|^{-\gamma_4})\right)g_3(\bk_0)t^{-d/2} + \epsilon(g_3)t^{-d/2-1}
 \end{equation}
 for $|\vec z|^2>\lambda_*$ and $0$ otherwise. Here,
 $$
 |\epsilon(g_3)| \leq c \sum_{{|m| \leq d+3}}\sup_{\left| \bk -\bk_0 \right| <2 } |D^{m}g_3(\bk)| \leq c\left\||\bk|^{d/2+2}\hatt \Psi_0(\bk)\right\|_{\mathcal C^{d+3}(\R^d)} \delta ^{-d-3}|\vec z|^{-d/2-1}.
 $$
 
Now, the end of the proof of Proposition~\ref{Prop2} follows as in Section 3 in \cite{KLSS}. The proof of Theorem~\ref{Thm1} is identical to the proof in Section 4 of \cite{KLSS}.

\begin{remark} \label{smoothEFE}

(a) The above proofs show that Theorem~\ref{Thm1} remains true if we replace $\mathcal C_0^{\infty}$ in (\ref{nontriv}) with
\[ {\mathcal S}_d := \{f:|x|^sD^{\bm}f(x) \in L^2(\R^d), 0\leq s,|m| \leq C(d)\}\]
 i.e.\ for initial conditions which are sufficiently smooth and of sufficiently rapid power decay. 

(b) Using the constructions in the above proofs, we can now also describe more explicitly how to choose initial conditions $\Psi_0$ for the solution of (\ref{IVP}) which give simultaneous ballistic upper and lower bounds. Essentially, one has to regularize elements in the range of $E_{\infty}$ in two different ways, first at the boundary of ${\mathcal G}_{\infty}$, using the cut-off function $\tilde\eta_{\delta}$ as in (\ref{eta-delta}) above, and then at high momentum $\bk$. For the latter, let $\varphi \in {\mathcal S}_d$ on $\R^d$  and such that $\varphi$ does not vanish identically on ${\mathcal G}_{\infty}$.

Choose
\begin{equation}
\Psi_0(\bx) :=\frac{1}{(2\pi)^{d/2}}\int_{\wti{\mathcalG}_\infty} \varphi(\bk) \,\tilde\eta_\delta(\bk) \,U_{\infty }(\bk,\bx) \,d\bk.
\end{equation}
As $\delta\to 0$ this converges to $F_0(\bx) = \frac{1}{(2\pi)^{d/2}} \int_{{\mathcal G}_{\infty}} \varphi(\bx) U_{\infty}(\bk, \bx)\, d\bk$ in the range of $E_{\infty}$ with $\|F_0\|^2 = \int_{{\mathcal G}_{\infty}} |\varphi|^2\, d{\bk}/(2\pi)^d\not=0$. Thus, for $\delta>0$ sufficiently small, $E_{\infty} \Psi_0 \not=0$.

Furthermore, our methods show that the choice of $\varphi \in {\mathcal S}_d$ gives $\Psi_0 \in {\mathcal S}_d$. Thus the initial condition $\Psi_0$ leads to a ballistic lower bound on transport. At the same time the condition of \cite{RaSi} for the ballistic upper bound (\ref{genballistic}) is satisfied.

\end{remark}


\begin{thebibliography}{100}

\bibitem{AK} J.~Asch and A.~Knauf,
{\em Motion in periodic potentials},
Nonlinearity 11 (1998), 175--200

\bibitem{AS} J.~Avron and B.~Simon, {\em Almost periodic Schr\"odinger operators, I. Limit periodic potentials}, Comm.\ Math.\ Phys.\ 82 (1981), 101--120


\bibitem{BSB} J.~Bellissard and H.~Schulz-Baldes, {\em Subdiffusive quantum transport for 3D Hamiltonians with absolutely continuous spectra}, J.\ Stat.\ Phys.\ 99 (2000), 587--594

\bibitem{B2007} J.~ Bourgain, {\em Anderson localization for quasi-periodic lattice Schr\"odinger operators on $\Z^d$, $d$ arbitrary}, Geom.\ Funct.\ Anal.\ 17 (2007), 682--706

\bibitem{BGS} J.~Bourgain, M.~Goldstein and W.~Schlag, {\em Anderson localization for Schr\"odinger operators on $\Z^2$ with quasi-periodic potential}, Acta Math.\ 188 (2002), 41--86

\bibitem{ChD} V.~A.~Chulaevsky and E.~I.~Dinaburg, {\em Methods of KAM-theory
for long-range quasi-periodic operators on $\Z^{\nu}$. Pure point spectrum}, Commun.\ Math.\ Phys.\ 153 (1993), 559--577

\bibitem{CD} V.~Chulaevsky and F.~Delyon, {\em Purely absolutely continuous spectrum for almost Mathieu operators}, J.\ Stat.\ Phys.\ 55 (1989), 1279--1284

\bibitem{C} J.-M.~Combes,
{\em Connections between quantum dynamics and spectral properties of time-evolution operators},
Differential equations with applications to mathematical physics,
Mathematics in Science and Engineering, 192 (1993), 59--68

\bibitem{DLS} D.~Damanik, D.~Lenz and G.~Stolz, {\em Lower transport bounds for one-dimensional continuum Schr\"odinger operators}, Math.\ Ann.\ 336 (2006), 361--389


\bibitem{DT} D.~Damanik and S.~Tcheremchantsev, {\em Power-law bounds on transfer matrices and quantum dynamics in one dimension}, Comm.\ Math.\ Phys.\ 236 (2003), 513--534

\bibitem{DT2} D.~Damanik and S.~Tcheremchantsev, {\em Scaling estimates for solutions and dynamical lower bounds on wavepacket spreading}, J.\ Anal.\ Math. 97 (2005), 103--131

\bibitem{DT3} D.~Damanik and S.~Tcheremchantsev, {\em Upper bounds in quantum dynamics}, J.\ Amer.\ Math.\ Soc.\ 20 (2007), 799-827

\bibitem{DT2010} D.~Damanik and S.~Tcheremchantsev, {\em A general description of quantum dynamical spreading over an orthonormal basis and applications to Schr\"odinger operators}, Discrete Contin.\ Dyn.\ Syst.\ 28 (2010), 1381--1412 

\bibitem{DS} E.~I.~Dinaburg and Ya.~Sinai, {\em The one-dimensional Schr\"odinger equation with a quasi-periodic potential}, Funct.\ Anal.\ Appl.\ 9 (1975), 279--289

\bibitem{El} L.~H.~Eliasson, {\em Floquet solutions for the 1-dimensional quasi-periodic Schr\"odinger equation}, Comm.\ Math.\ Phys.\ 146 (1992), 447--482

\bibitem{Ge} I.~M.~Gel'fand,
{\em Expansion in Eigenfunctions of an Equation with Periodic Coefficients.}
Dokl. Akad. Nauk SSSR, 73 (1950), 1117--1120 (in Russian).

\bibitem{GKT} F.~Germinet, A.~Kiselev and S.~Tcheremchantsev,
{\em Transfer matrices and transport for Schr\"{o}dinger operators},
Ann.\ Inst.\ Fourier 54 (2004), 787--830


\bibitem{G} I.~Guarneri,
{\em Spectral properties of quantum diffusion on discrete lattices},
Europhysics Lett.\ 10 (1989), 95--100

\bibitem{G2} I.~Guarneri, {\em On an estimate concerning quantum diffusion in the presence of a fractional spectrum}, Europhys.\ Lett.\ 21 (1993), 729--733


\bibitem{H1} L.~H\"ormander,
{\it The analysis of linear partial differential operators. I. Distribution theory and Fourier analysis},
Springer, 256 (1990)

\bibitem{JSBS} S.~Jitomirskaya, H.~Schulz-Baldes and G.~Stolz, {\em Delocalization in random polymer models}, Comm.\ Math.\ Phys.\ 233 (2003), 27--48

\bibitem{JM} R.~Johnson and J.~Moser, {\em The rotation number for almost periodic potentials}, Comm.\ Math.\ Phys.\ 84 (1982), 403--438

\bibitem{KL1} Yu.~Karpeshina and Y.-R.~Lee,
{\em Spectral properties of polyharmonic operators with limit-periodic potential in dimension two},
J.\ Anal.\ Math.\ 102 (2007), 225--310

\bibitem{KL2} Yu.~Karpeshina and Y.-R.~Lee,
{\em Absolutely continuous spectrum of a polyharmonic operator with a limit-periodic potential in dimension two},
Comm.\ Partial Differential Equations 33 (2008), 1711--1728

\bibitem{KL3} Yu.~Karpeshina and Y.-R.~Lee,
{\em Spectral properties of a limit-periodic Schr\"odinger operator in dimension two},
J.\ Anal.\ Math.\ 120 (2013), 1--84

\bibitem{KLSS} Yu.~Karpeshina, Y.-R.~Lee, R.~Shterenberg, G.~Stolz, {\em Ballistic transport for the Schr\"odinger operator with limit-periodic or quasi-periodic potential in dimension two}, CMP, August 2017, Vol. 354, no. 1, pp 85 -- 113.

\bibitem{KPS} Yu.~Karpeshina, L.~Parnovski, R.~Shterenberg, {\em Bethe-Sommerfeld Conjecture and Absolutely Continuous Spectrum of Multi-Dimensional Quasi-Periodic Schr\"odinger Operators}, arxiv:2010.05881 

\bibitem{KS} Yu.~Karpeshina and R.~Shterenberg,
{\em Multiscale analysis in momentum space for quasi-periodic potential in dimension two},
J.\ Math.\ Phys.\ 54, 073507 (2013), 1--92

\bibitem{KS1} Yu.~Karpeshina and R.~Shterenberg, {\em Extended States for the Schr\"odinger Operator with Quasi-periodic Potential in Dimension Two}, Memoirs of AMS, {\bf 258},
\# 1239 (third of 7 numbers), 2019.

\bibitem{KL} A.~Kiselev and Y.~Last,
{\em Solutions, spectrum, and dynamics for Schr\"odinger operators on infinite domains},
Duke Math.\ J.\ 102 (2000), 125--150

\bibitem{L} Y.~Last,
{\em Quantum dynamics and decompositions of singular continuous spectra},
J.\ Funct.\ Anal.\ 142 (1996), 406--445

\bibitem{MC} S.~A.~Molchanov and V.~Chulaevsky, {\em The structure of a spectrum of lacunary-limit-periodic Schr\"odinger operator}, Functional Anal.\ Appl.\ 18 (1984), 343--344

\bibitem{MP} J.~Moser and J.~P\"oschel, {\em An extension of a result by Dinaburg and Sinai on quasiperiodic potentials}, Comment.\ Math.\ Helv.\ 59 (1984), 39--85

\bibitem{PF} L.~Pastur and A.~Figotin, {\em Spectra of random and almost-periodic operators}, Springer, 1992

\bibitem{RaSi} C.~Radin and B.~Simon,
{\em Invariant domains for the time-dependent Schr\"odinger equation},
J.\ Differential Equations 29 (1978), 289--296

\bibitem{ReedSimon}  M.~Reed and B.~Simon, {\em Methods of modern mathematical physics. II. Fourier analysis, self-adjointness}. Academic Press, New York-London, 1975

\bibitem{R} H.~R\"ussmann, On the one dimensional Schr\"odinger equation with quasi-periodic potential, Ann.\ N.\ Y.\ Acad.\ Sci.\ 357 (1980), 90--107

\bibitem{SkSo}  M.~M.~ Skriganov and A.~V.~Sobolev,
{\em On the spectrum of a limit-periodic Schr\"odinger operator},
Algebra i Analiz, 17 (2005), 5;
English translation: St.\ Petersburg Math.\ J.\ 17 (2006), 815--833

\bibitem{Tch} S.~Tcheremchantsev,
{\em Mixed lower bounds for quantum transport},
J.\ Funct.\ Anal.\ 197 (2003) 247--282

\end{thebibliography}
\end{document}